\documentclass[aps,%
               prl,%
               reprint,%
               showpacs,%
               superscriptaddress,%
               longbibliography,%
               nofootinbib%
               ]{revtex4-2}

\usepackage[utf8]{inputenc}
\usepackage[usenames,dvipsnames]{xcolor}
\usepackage{graphicx}
\usepackage{amsfonts}
\usepackage{amssymb}
\usepackage{amsmath}
\usepackage{mathtools}
\usepackage{bm}
\usepackage{dsfont}
\usepackage{braket}
\usepackage{tikz}
\usepackage{txfonts}

\usepackage[pdfusetitle,%
            bookmarks=true,%
            colorlinks,%
            linkcolor=blue,%
            citecolor=blue,%
            urlcolor=blue%
            ]{hyperref}

\newcommand{\eq}[1]{Eq.\thinspace(\ref{#1})}

\newcommand{\subfigref}[2]{\hyperref[fig:#1]{\ref*{fig:#1}(#2)}}

\begin{document}

\def\papertitle{{Altermagnetism of Ultracold Atoms in Optical Lattices}}
\def\papertitle{{Fermi-Hubbard Altermagnetism of Ultracold Atoms}}
\def\papertitle{{Realizing Altermagnetism in Fermi-Hubbard Models with Ultracold Atoms}}

\def\tum{{Technical University of Munich, TUM School of Natural Sciences, Physics Department, 85748 Garching, Germany}}
\def\mcqst{{Munich Center for Quantum Science and Technology (MCQST), Schellingstr. 4, 80799 M{\"u}nchen, Germany}}
\newcommand{\TUM}{\affiliation{\tum}}
\newcommand{\MCQST}{\affiliation{\mcqst}}

\title{\papertitle}
\author{Purnendu Das} \TUM \MCQST  \affiliation{Indian Institute of Science, Bangalore, 560012, India} 
\author{Valentin Leeb} \TUM \MCQST
\author{Johannes Knolle} \TUM \MCQST 
\affiliation{Blackett Laboratory, Imperial College London, London SW7 2AZ, United Kingdom}
\author{Michael Knap} \TUM \MCQST

\date{\today}

\begin{abstract}
Altermagnetism represents a type of collinear magnetism, that is in some aspects distinct from ferromagnetism and from conventional antiferromagnetism. In contrast to the latter, sublattices of opposite spin are related by spatial rotations and not only by translations and inversions. As a result, altermagnets have spin-split bands leading to unique experimental signatures. Here, we show theoretically how a $d$-wave altermagnetic phase can be realized with ultracold fermionic atoms in optical lattices. 
We propose an altermagnetic Hubbard model with anisotropic next-nearest neighbor hopping and obtain the Hartree-Fock phase diagram. The altermagnetic phase separates in a metallic and an insulating phase and is robust over a large parameter regime. We show that one of the defining characteristics of altermagnetism, the anisotropic spin transport, can be probed with trap-expansion experiments.

\end{abstract}

\maketitle

\textbf{\textit{Introduction.---}}Collinear quantum magnets are usually assumed to have either ferromagnetic or antiferromagnetic order~\cite{ashcroft2022solid,auerbach1998interacting}. Ferromagnets break time-reversal symmetry leading to spin-split bands and a net polarization of the magnetic moment. Conventional antiferromagnets exhibit zero net magnetization and are symmetric under translation and spin-inversion, leading to spin-degenerate bands.
However, recent studies have suggested refinements of this dichotomy and proposed a new class of collinear magnetism , that possess momentum dependent spin-split bands without net magnetization~\cite{wu2007fermi,ahn2019antiferromagnetism,Hayami_2019,Smejkal2020crystal,Hayami_2020,mazin2021prediction,Smejkal_2022_b,Shao_2021,smejkal2022giant,Smejkal_2022_a,Mazin2022editorial,Maier_2023},
as recently confirmed experimentally in material candidates~\cite{krempasky2023altermagnetic,lee2023broken,reimers2023direct,Feng2022,Zhou2024}. These collinear states, dubbed  altermagnets~\cite{Smejkal_2022_b,Smejkal_2022_a,Mazin2022editorial},
are characterized by a rotational symmetry of the opposite spin sublattices. For example, in a $d$-wave altermagnet on the square lattice, sublattices are related by a spin flip followed by a $\pi/2$ real-space rotation about a point on the dual lattice; see Fig.~\ref{fig:alter}~(a) for an illustration.

\begin{figure}
	\centering
	\includegraphics[width=1\linewidth]{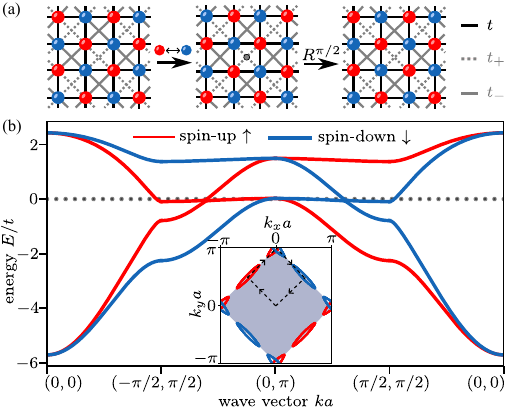}
	\caption{\textbf{The altermagnetic Hubbard model.} (a) A N\'{e}el state on an alternating anisotropic square lattice is a $d$-wave  altermagnetic state. It is invariant under a global spin flip (exchanging red and blue dots) followed by a $\pi/2$ real-space rotation around the dual square lattice (gray dot). The nearest-neighbor hopping $t$ and the alternating diagonal hopping $t_\pm$ of the altermagnetic Hubbard model are indicated as well. (b) Spin-resolved band structure of the altermagnetic state at zero temperature for $t'/t = 0.3, \delta = 0.9$, $U/t = 3.5$ evaluated along the path indicated in the inset. Inset: Fermi surface and magnetic Brillouin zone (gray shaded area). The band structure obeys the symmetry shown in (a) and is therefore spin-split without net magnetization, which are the key characteristics of an altermagnet.}
	\label{fig:alter}
\end{figure}

Over the recent years, exciting progress has been made in studying quantum magnetism with quantum simulators of ultracold atoms~\cite{Bloch_2008}. For the square lattice Hubbard model antiferromagnetic correlations of an extended range have been observed at the lowest experimentally accessible temperatures~\cite{Jordens2008,Schneider2008,Hart2015,Mazurenko2017} and the consequences of doping the antiferromagnetic state have been investigated~\cite{Chiu2019,Bohrdt2019,Koepsell2021,Bohrdt2021}. %Frustrated triangular lattice Hubbard models have been started to be explored as well~\cite{Mongkolkiattichai2022, Xu2023, Lebrat2023, Prichard2023}. 
Investigating the phenomena of altermagnetism with ultracold atoms remains an interesting open avenue. 

In this work, we show how $d$-wave altermagnetism can be realized and characterized with ultracold atoms in optical lattices. We analyze a square lattice Hubbard model with uniform nearest-neighbor and alternating diagonal hoppings and show how this model can be realized by $45^\circ$ rotated optical lattices. Performing a Hartree-Fock analysis we find that this model stabilizes an altermagnetic phase in an extended parameter range and analyze the robustness of the state at finite temperatures. We demonstrate that the key experimental characteristic of the altermagnetic state, i.e., the anisotropic spin transport, can be measured by trap-expansion experiments.

\textbf{\textit{The altermagnetic Hubbard model.---}}We consider two-species of fermionic atoms labelled by spin $s$ in an optical lattice described by the following altermagnetic Hubbard model
\begin{equation}
  \hat H = - \sum_{i,j, s} t_{ij} (c_{i s}^\dagger c_{j s} + \text{h.c}) + U \sum_{i} n_{i \uparrow}n_{i \downarrow},
\label{eq:Hamiltonian}
\end{equation}
where $U$ is the on-site Hubbard interaction and $t_{ij}$ the hopping matrix element, which is uniform and of strength $t$ for nearest neighbors, sublattice-dependent for diagonal neighbors, and zero otherwise. 
The diagonal hopping alternates with $t_\pm = t'(1\pm \delta)$ as following: in the $(1,1)$-direction the hopping element is $t_-$ ($t_+$) and in the  $(1,-1)$-direction it is $t_+$ ($t_-$) on the $A$ (B) sublattice, respectively; see Fig.~\ref{fig:alter}~(a).

We consider half-filling $\langle n_i \rangle = \langle n_{i\uparrow} + n_{i\downarrow} \rangle = 1$. However, our results remain qualitatively similar for small doping where the N\'{e}el order is stable. We will now show that this particular sublattice dependence of the diagonal hopping leads to altermagnetism and discuss later the optical lattice geometry required to realize this model. 

In order to study the magnetic instabilities of our system, we perform a Hartree-Fock analysis that captures the sublattice structure of the $(\pi,\pi)$ magnetic instability. To this end, we introduce the altermagnetic order parameter, $\delta m = \frac{1}{4N}\sum_r \left<n_{rA\uparrow}-n_{rA\downarrow}-n_{rB\uparrow} +n_{rB\downarrow}\right>$, which is proportional to the staggered magnetization. At filling $n$ we write the occupation
\begin{equation}
\langle n_{r \lambda s} \rangle = n/2 + \delta m (-1)^{\lambda+s},
\label{eq:4}
\end{equation}
where $r$ denotes the index of a unit cell, $\lambda$ the sublattice, and $s$ the spin. For the alternating sign of the order parameter $(-1)^{\lambda+s}$ we associate $\lambda$ and $s$ with 0 for A and $\uparrow$ and with 1 for B and $\downarrow$, respectively. A non-zero order parameter $\delta m$ indicates a sublattice Néel ordering which in conjunction with the lattice symmetries gives rise to the altermagnetic state. Decoupling the interaction term and expressing it in terms of the mean-field order parameter leads after Fourier transformation to the effective interactions $-U\delta m \sum_{\bm{k}} (n_{\bm{k} A \uparrow} -n_{\bm{k} B \uparrow} - n_{\bm{k} A \downarrow} +n_{\bm{k} B \downarrow})$, where the wave vector $\bm{k}$ is in the magnetic Brillouin zone. The magnetic Brillouin zone is defined via the real space unit cell spanned by the primitive vectors $\bm{a}_1 = a(1,1)$ and $\bm{a}_2 = a(1,-1)$ with the lattice constant $a$ of the square lattice.

Expressing the mean-field Hamiltonian in the basis of $\Psi_{\bm{k}}^\dagger = (c_{{\bm{k}} A \uparrow}^\dagger, c_{{\bm{k}} B \uparrow}^\dagger, c_{{\bm{k}} A \downarrow}^\dagger, c_{{\bm{k}} B \downarrow}^\dagger)$, leads to
\begin{equation}
\begin{split}
  \hat{H}^\text{HF}  = \sum_{\bm{k}}\Psi_{\bm{k}}^\dagger \vcenter{\hbox{$\begin{bmatrix}
    H_{\uparrow}(\bm{k}) & 0 \\
    0 & H_{\downarrow}(\bm{k})\\
\end{bmatrix}$}} \Psi_{\bm{k}}.   
\end{split}
	\label{equ:7}
\end{equation}
The Hamiltonian is block diagonal in the spin degree of freedom with $H_{s}(\bm{k}) = \vcenter{\hbox{$\begin{bmatrix}
    h_{AA,s} & h_{AB,s} \\
    h_{BA,s} & h_{BB,s}\\
\end{bmatrix}$}}$, 
where $h_{AA,s} = -[2t_-\cos(\bm{k} \bm{a}_1)+2t_+\cos(\bm{k} \bm{a}_2) + (-1)^s U\delta m]$, $h_{BB,s} = -[2t_+\cos(\bm{k} \bm{a}_1)+2t_-\cos(\bm{k} \bm{a}_2) - (-1)^s U\delta m]$, $h_{AB,s} = -2t[\cos(k_xa)+\cos(k_ya)]$, and $h_{BA,s} = h_{AB,s}^*$. Due to the spin block-diagonal structure of the Hamiltonian \eqref{equ:7}, bands are fully spin-polarized. In addition, each of the spin components exhibit a momentum-inversion symmetry ($\bm{k} \to -\bm{k}$) and sublattices are staggered.

We solve the mean-field equations at finite temperatures $T$ by self-consistently determining the order parameter $\delta m = \frac{1}{4N} \sum_{\bm{k}} \langle n_{\bm{k}A \uparrow} -n_{\bm{k}B \uparrow} - n_{\bm{k}A \downarrow} +n_{\bm{k}B \downarrow}\rangle_\text{HF}$ as well as the chemical potential $\mu$, which is set by fixing the particle number; see supplemental materials for details~\cite{supp}. We compute the spin-resolved band structure for $t'/t = 0.3, \delta = 0.9$, and $U/t = 3.5$ at half-filling $n = 1$; see Fig.~\ref{fig:alter}~(b). Both the band structure and the Fermi surface possess the altermagnetic symmetry of a $\pi/2$ rotation along with a spin-flip. Here, the reciprocal lattice vectors of the magnetic Brillouin zone are $\frac{\pi}{2a}(1,1) $ and $\frac{\pi}{2a}(1,-1) $; see shaded area in the inset of Fig.\ref{fig:alter} (b).

\begin{figure}
	\centering
	\includegraphics[width=1\linewidth]{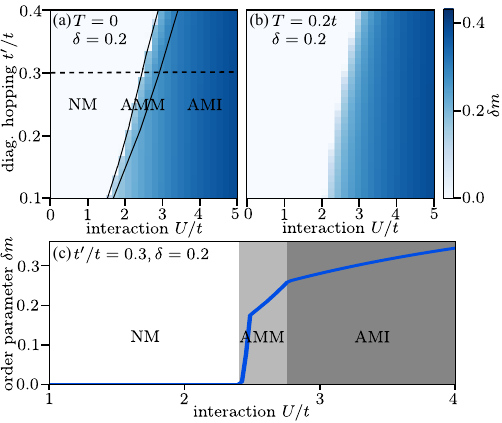}
	\caption{\textbf{Robustness of altermagnetism.} We compute the altermagnetic order parameter $\delta m$ as a function of interaction strength $U/t$ and diagonal hopping $t'/t$ for staggering $\delta = 0.2$ and temperatures (a) $T=0$ and (b) $T=0.2t$. The system is in a normal metallic state when the order parameter vanishes, while a finite order parameter indicates altermagnetic symmetry breaking.     
   (c) Line cut along the dashed line in (a) shows the order parameter at zero temperature for $t'/t = 0.3$ and $\delta = 0.2$. Three phases are distinguished:  The normal metal (NM) at weak interactions, the altermagnetic metal (AMM) possessing a Fermi surface at intermediate interactions, and the gapped altermagnetic insulator (AMI) at strong interactions. Within the AMM the kink in the order parameter at $U/t \approx 2.5$ indicates a Lifshitz transition at which half of the Fermi pockets vanish.} 
	\label{fig:PhaseDiagram}
\end{figure}

Having established the altermagnetic state, we study its robustness by tuning the system parameters and the temperature. To this end, we compute the order parameter $\delta m$ as a function of $U/t$ and $t'/t$ for $\delta = 0.2$ and $T=0$ and $0.2t$ in Fig.~\ref{fig:PhaseDiagram}~(a,b). We will show below that the hopping parameters can be controlled  by the optical lattice. Moreover, the interaction $U$ is tunable by Feshbach resonances in ultracold atomic systems~\cite{Chin2010}. The altermagnetic phase is stabilized for increasing diagonal hopping $t'$, staggering $\delta$, and interaction strength $U$. It can be either metallic, characterized by the presence of small Fermi surfaces, or a gapped insulator. 
A line cut though the phase diagram unveils phase transitions from a normal metal (NM) with vanishing $\delta m$ over an altermagnetic metal (AMM) to an altermagnetic insulator (AMI); see Fig.~\ref{fig:PhaseDiagram}~(c). In addition, we find a kink in the order parameter $\delta m$ within the AMM at $U/t \approx 2.5$. This is a Lifshitz transition at which half of the Fermi pockets around $(\pm\pi/2a,\pm\pi/2a)$ disappear. The second kink then indicates the transition from AMM to AMI phase at which the Fermi
surfaces disappear. The altermagnetic phase occupies a large portion of the phase diagram, because the underlying mechanism is a consequence of the symmetry of the single-particle band structure. The interactions are only required to establish N\'{e}el order, which splits the bands appropriately.
The spin-splitting is set by the anisotropy of the next-nearest neighbor hopping  $t' \delta$. When decreasing  $t' \delta$ to zero, the spin-flip and $C_4$-rotation symmetry of the altermagnet regains the spin-flip and translation symmetry, and a conventional   antiferromagnet is realized.

\begin{figure}
	\centering
	\includegraphics[width=1\linewidth]{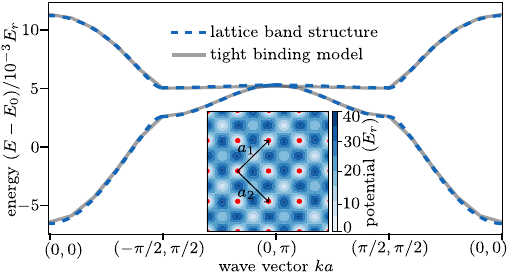}
	\caption{\textbf{Effective band structure of the optical lattice.} Lowest two bands, solid lines, obtained from solving the Schr\"dinger equation of a particle in an optical lattice potential with with $V_0 = 4 E_r, V_1/V_0 = 2.2$, and $\Delta = 0.6$; see Eq. \eqref{eq:1}. Bands obtained from a tight-binding model with uniform nearest-neighbor hopping $t$ and the alternating diagonal hopping amplitudes $t_\pm = t'(1\pm\delta)$, dashed line, agree well with the full band structure. The effective parameters of the tight-binding model are $t = 2.2 \times 10^{-3} E_r, t'/t = 0.16, \delta = 0.83$ and the energy offset is $E_0 = 13.81 E_r$. Inset: Illustration of the optical lattice potential. 
 }
	\label{fig:OpticalLattice}
\end{figure}
\textbf{\textit{Optical lattice for the altermagnetic band structure.---}}The altermagnetic Hubbard model has uniform nearest-neighbor and alternating diagonal  hopping elements. Such a single-particle band structure is realized when considering $45^\circ$-rotated counter-propagating and phase-locked lasers of wave length $\lambda$ and $2\lambda$ with different strengths, respectively. Specifically, we consider the following lattice potentials
\begin{equation}
\begin{split}
V_\text{latt} &= E_r(V_\text{sq} + V_{d,1} + V_{d,2}) \\
V_\text{sq} &=  V_0  [\sin^2 (k_l x) + \sin^2 (k_l y)]\\
V_{d,1} &= V_1  \Big[\Delta_+ \sin^2 [k_l (x+y)]+\Delta_- \sin^2 [\frac{k_l}{2} (x+y)]\Big]\\
V_{d,2} &= V_1 \Big[ \Delta_+ \sin^2 [k_l (x-y)] + \Delta_- \cos^2 [\frac{k_l}{2} (x-y)]\Big],
\end{split}
\label{eq:1}
\end{equation}
where $k_l = 2\pi / \lambda$ is the lattice wave vector, $E_r = \frac{\hbar^2k_l^2}{2m}$ is the recoil energy, $m$ is the mass of the atoms, and $\Delta_\pm = (1\pm\Delta)$ the potential staggering of strength $\Delta$. The potential consists of deep minima on a square lattice and shallow minima on the dual lattice that are tuned by  $V_0$, $V_1$, and $\Delta$; see inset of Fig.~\ref{fig:OpticalLattice}. For such an optical lattice both the nearest neighbor and the diagonal tunneling are sizeable. 

The unit cell of the lattice is $\sqrt{2} a\times \sqrt{2}a$, where $a$ is the lattice constant of the square lattice, with primitive lattice vectors $\bm{a}_1 $ and $\bm{a}_2$; see inset of Fig. \ref{fig:OpticalLattice}.
We numerically solve the Schr\"odinger equation of a single particle in this optical lattice potential  by standard techniques (see e.g. Ref.~\onlinecite{Bissbort_2013}) and show the lowest two bands in Fig.~\ref{fig:OpticalLattice}. We then fit the lowest bands to the tight-binding Hamiltonian of the altermagnetic Hubbard model and obtain the uniform nearest-neighbor hopping $t$ and the staggered diagonal hoppings $t_\pm = t'(1\pm \delta)$. The diagonal hopping elements are sizeable for this lattice because of the potential minima at the dual lattice sites. The tight-binding band structure reproduces well the lowest two bands; Fig.~\ref{fig:OpticalLattice}. The tight-binding parameters are tunable by $V_0$, $V_1$, and $\Delta$ which characterize the optical lattice; see supplemental materials~\cite{supp}. Here, we have considered deep optical lattices, leading to comparatively low absolute scales of the hopping. Shallower lattices provide larger absolute hopping scales, while at the same time giving rise to longer ranged hoppings. As a consequence more complex tight-binding models are needed for quantitative agreement. However, we find that the band structure of shallower lattice potentials still posses the same symmetries and the same anisotropic behavior as 
the deep lattice; see supplemental materials~\cite{supp}. Larger effective energy scales will be advantageous in experiments to access the required temperatures and obtain homogeneous parameters throughout the system.

\textbf{\textit{Experimental signatures.---}}The altermagnetic state manifests itself in a vanishing net magnetization but has a pronounced spin-polarized Fermi surface, which can be probed by spin-resolved transport \cite{Smejkal2020crystal,Gonzalez2021_efficient}. One way to probe such anomalous transport with ultracold atoms is to release the trapping potential and to subsequently measure the spin-resolved densities while the atomic cloud expands, see e.g. Refs.~\cite{Sommer2011, Schneider2012,Brown2019,Nichols2019,Oppong2022}. To characterize such an expansion experiment, we first determine the conductivity tensor and then use Einstein's relation for the diffusion constant to obtain an effective hydrodynamic description of the expansion dynamics. 

The conductivity tensor for both spin-up and spin-down atoms are $2\times 2$ matrices with elements $\sigma^s_{\alpha \beta}$, where $\alpha, \beta \in \{\tilde x, \tilde y\}$ indicate the spatial direction along the primitive lattice vectors $\{\bm{a}_1$, $\bm{a}_2\}$ of the two-site unit cell and $s\in \{\uparrow, \downarrow\}$ is the spin state. Since the bands are fully spin-polarized, the conductivity is diagonal in spin basis, see  \eq{equ:7}. The transverse Hall contribution to the conductivity vanishes, $\sigma_{\alpha\beta}^s=0$ for $\alpha \neq \beta$, due to the momentum-inversion symmetry of \eq{equ:7}. From the spin-flip and $\pi/2$ rotation symmetry in real space, we further deduce that the conductivity tensor of spin-up and spin-down are  related by $\sigma^{\uparrow}_{\alpha\alpha} = \sigma^{\downarrow}_{\bar\alpha,\bar \alpha}$, where $\bar \alpha$ is the direction orthogonal to $\alpha$. We use the Kubo formula to evaluate the diagonal DC conductivity tensor~\cite{Kubo_1957,Crepieux_2001,Freimuth_2014}, see also supplemental materials~\cite{supp}

\begin{align}
    \sigma_{\alpha\alpha}^s = -\frac{\hbar}{\pi V} &\int_{-\infty}^{\infty} d\epsilon \frac{df}{d\epsilon} \sum_{m,n,\bm{k}} |\langle \psi_m(\bm{k})| v_\alpha^s |\psi_n(\bm{k})\rangle|^2\nonumber\\
    &\times  \frac{\Gamma}{(\epsilon - \epsilon_n)^2+\Gamma^2} \frac{\Gamma}{(\epsilon - \epsilon_m)^2+\Gamma^2},
\end{align}
where ${v}^s_\alpha = \frac{1}{\hbar}\nabla_{{k_\alpha}} H_s(\bm{k})$ is the spin dependent velocity, $f$ is the Fermi-Dirac distribution function, $\epsilon_n$ and $|\psi_n(\bm{k})\rangle$ are the eigenenergies and -states of \eq{equ:7}, respectively, $\Gamma$ is an positive infinitesimal that we use for the numerical evaluation of the integral.

In order to compute the relaxation dynamics, we relate the conductivity matrix with the diffusion matrix by the Einstein relation \cite{Kubo_1957},
\begin{equation}
 \sigma^s_{\alpha \beta} = \frac{n^s D_{\alpha \beta}^s}{T},
 \label{eq:eins.rel}
\end{equation}
where $n^s$ is the particle density of spin $s$ atoms and $T$ is the temperature. Our model conserves the densities of both spin species separately, leading to the continuity equation  $\frac{\partial n^s}{\partial \tau}  + \nabla  \bm{J}^s = 0$, where $\tau$ denotes real time. Taking the hydrodynamic assumption, we perform a gradient expansion of the currents. Due to the symmetries of the conductivity tensor, only diagonal contributions arise and the currents are related to the density gradients as $J_\alpha^s = - D_{\alpha\alpha} \partial_\alpha n^s$, where $\alpha \in \{\tilde x, \tilde y\}$. We thus obtain the  diffusion equation
\begin{equation}
 \frac{\partial n^s}{\partial \tau}  =  (D^s_{\tilde x \tilde x} \partial^2_{\tilde x}  + D^s_{\tilde y \tilde y} \partial^2_{\tilde y})n^s.
\label{eq:16}
\end{equation}
As the diffusion constants are anisotropic in space, the transport of spin will be anisotropic as well. This is a key signature of the altermagnetic state. In order to demonstrate this behavior, we initialize our system at temperature $T = 0.2 t$ in the optical potential characterized by $V_0 = 4 E_r, V_1/V_0 = 2.2 E_r,\Delta = 0.6$ and $\lambda = 1064$\,nm at half-filling inside the square-shaped region;  for the temperature dependence of the conductivity see supplemental material~\cite{supp}. Subsequently, we let the particles expand by removing the trapping potential at time $\tau = 0$ and compute the time evolution of the spin-resolved densities by numerically solving the diffusion equation~\eqref{eq:16}; Fig.~\ref{fig:Diffusion}~(a).

\begin{figure}
	\centering
	\includegraphics[width=1\linewidth]{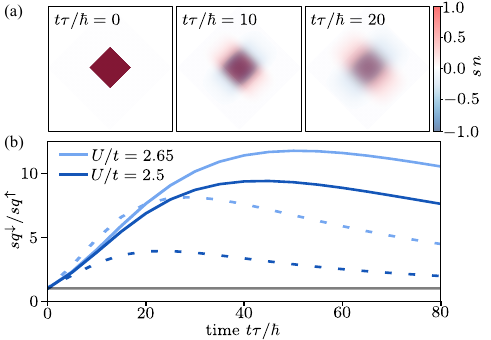}
	\caption{\textbf{Anisotropic spin diffusion.} (a) Trap-release dynamics of an altermagnetic  state at finite temperature $T=0.2 t$, trapped in a box potential in an optical lattice characterized by $V_0 = 4 E_r, V_1/V_0 = 2.2,\Delta = 0.6$. The spin-resolved density propagates anisotropically in real space along the $\bm{a}_1$ and $\bm{a}_2$ directions. %Here $n$ denotes the density of atoms and $s$ is the spin of the atom. 
    (b) We characterize the anisotropic expansion by the ratio of the geometric squeezing parameter ${sq}^s(\tau)$ of spin-down and spin-up atoms for two different values of the interaction $U$ and two different temperatures  $T = 0.15t$ (solid lines) and $T = 0.2t$ (dashed lines). The grey line represents the isotropic expansion of a normal metallic state for which ${sq}^s(\tau)$ is always one.}
	\label{fig:Diffusion}
\end{figure}

We observe that the spin-up and spin-down atoms predominantly relax in different directions, related by a $\pi/2$ real-space rotation. The spin-up atoms have a larger contribution to $\sigma^\uparrow_{\tilde y \tilde y}$ than $\sigma^\uparrow_{\tilde x \tilde x}$ as can be also seen from the Fermi surface in the inset of Fig.~\ref{fig:alter}~(b). Thus diffusion is stronger in the $\bm{a}_2$-direction than in $\bm{a}_1$ direction and vice versa for spin-down atoms. To quantify the anisotropy, we define a geometric squeezing parameter
\begin{equation}
 {sq}^s(\tau) = \frac{\int d^2  \tilde r \,\tilde x^2 \,n^s(\bm{\tilde r},\tau)}{\int d^2 \tilde r \,\tilde y^2 \,n^s(\bm{\tilde r},\tau) },
 \label{eq:17}
\end{equation}
which measures the relative spread in $\tilde x$-direction compared to the $\tilde y$-direction. The relative squeezing of spin-down and spin-up $sq^{\downarrow}(\tau)/sq^{\uparrow}(\tau)$ initially increases strongly and then approaches one asymptotically because the steady state is uniform in space; Fig.~\ref{fig:Diffusion}~(b). 
 
When increasing the interaction strength $U$ the altermagnetic order parameter increases and by consequence also the spin-splitting energy, which leads to a larger squeezing parameter. 
For higher temperatures the anisotropy in the conductivity tensor is reduced as the spin splitting decreases. However, the initial growth of $sq^{\downarrow}(\tau)/sq^{\uparrow}(\tau)$ can be larger as overall the diffusion constant increases with temperature according to the Einstein relation.  Although we assume the experiment to be performed in a box potential, one could also release a harmonic confinement potential~\cite{Oppong2022}. The central idea is to observe the anisotropic spin diffusion which is a characteristic of the altermagnetic phase.

\textit{\textbf{Conclusions \& Outlook.---}}Altermagnetism represents a type of collinear magnetism, that is characterized by rotational symmetries between opposite spin sublattices. We have shown how such an altermagnetic state can be realized with fermionic ultracold atoms. As the underlying mechanism derives from the single-particle band structure, the state is robust and arises over a large parameter range. We discuss that the unconventional symmetry of the state can be detected experimentally in  trap expansion experiments which exhibit anisotropic expansion for the different spin species.

Signatures of the altermagnetic state can also be obtained with quantum gas microscopes~\cite{Bakr_2009, Sherson_2010}. The order parameter, i.e., the staggered magnetization, can be measured directly as a real space N\'eel pattern~\cite{Mazurenko2017}. In addition, anisotropic local current distributions need to be established for altermagnetism. The rung current can be measured in quantum gas microscopes, by freezing the states into double wells and subsequently performing a $\pi/4$ tunneling event in each double well. This
maps the rung current to the occupation basis~\cite{Wybo2023} from which the altermagnetic symmetries can be deduced.

Our work demonstrates the potential for ultracold atoms to provide a controllable platform for realizing and probing   this new form of magnetism and for understanding the structure of fluctuations around the ordered states. 
For future work it would be interesting to characterize the anisotropic spin-susceptibilities of the altermagnetic state, which can be measured for example by Ramsey interferometry~\cite{Knap2013} or modulation spectroscopy~\cite{Bohrdt2018}. Furthermore, the real-time dynamics of spin-wave excitations in the altermagnetic insulating state could unveil the unconventional symmetry of the state as well. An exciting direction is to explore the interplay of doped altermagnets and  competing superconducting instabilities, which may offer a route to realize finite-momentum pairing or topological superconductivity.

\textbf{\textit{Acknowledgements.---}} 
We acknowledge support from the Deutsche Forschungsgemeinschaft (DFG, German Research Foundation) under Germany’s Excellence Strategy--EXC--2111--390814868 and DFG Grants No. KN1254/1-2, KN1254/2-1,  TRR 360 - 492547816, the European Research Council (ERC) under the European Union’s Horizon 2020 research and innovation programme (Grant Agreement No. 851161), as well as the Munich Quantum Valley, which is supported by the Bavarian state government with funds from the Hightech Agenda Bayern Plus. J.K. acknowledges support from the Imperial-TUM flagship partnership. V.L. acknowledges support from the Studienstiftung des deutschen Volkes. P.D. acknowledges support from the Working Internship in Science and Engineering (WISE) from the Deutscher Akademischer Austauschdienst (DAAD).

\textbf{\emph{Data and Code availability.---}}Numerical data and simulation codes are available on Zenodo~\cite{zenodo}.

\bibliography{reference}

\newpage
\leavevmode \newpage

\setcounter{equation}{0}
\setcounter{page}{1}
\setcounter{figure}{0}
\renewcommand{\thepage}{S\arabic{page}}  
\renewcommand{\thefigure}{S\arabic{figure}}
\renewcommand{\theequation}{S\arabic{equation}}
\onecolumngrid
\begin{center}
\textbf{Supplemental Material:}\\
\textbf{\papertitle}\\ \vspace{10pt}
Purnendu Das$^{1,2,3}$, Valentin Leeb$^{1,2}$, Johannes Knolle$^{1,2,4}$, and Michael Knap$^{1,2}$ \\ \vspace{6pt}

$^1$\textit{\small{Technical University of Munich, TUM School of Natural Sciences, Physics Department, 85748 Garching, Germany}} \\
$^2$\textit{\small{Munich Center for Quantum Science and Technology (MCQST), Schellingstr. 4, 80799 M{\"u}nchen, Germany}} \\
$^3$\textit{\small{Indian Institute of Science, Bangalore, 560012, India}}\\
$^4$\textit{\small{Blackett Laboratory, Imperial College London, London SW7 2AZ, United Kingdom}} \\
\vspace{10pt}
\end{center}
\maketitle
\twocolumngrid

\subsection{Self-consistent Hartree-Fock Equations}

We analyze the altermagnetic instabilities of our system within Hartree-Fock theory in which we decouple the interactions as $U  n_{r\lambda \uparrow}n_{r\lambda \downarrow} \approx\hspace{2mm} U  n_{r\lambda \uparrow} \left<n_{r\lambda \downarrow}\right> +  U  \left<n_{r\lambda \uparrow}\right>n_{r\lambda \downarrow} - U \left<n_{r\lambda\uparrow}\right>\left<n_{r\lambda \downarrow}\right>$, where $r$ is the index of a unit cell and $\lambda$ denotes the sub-lattice. Rewriting the interactions with the order parameter $\delta m$, defined in \eq{eq:4}, we obtain
\begin{equation}
\begin{split}
U\sum_{r,\lambda} n_{r\lambda \uparrow}&n_{r\lambda \downarrow} \approx \sum_{r,\lambda} - U\delta m (-1)^{\lambda}(n_{r\lambda \uparrow} - n_{r\lambda \downarrow})\\
&+U\frac{n}{2}(n_{r\lambda \uparrow}+n_{r\lambda \downarrow})
-U\left(\frac{n^2}{4} - \delta m^2\right).
\end{split}
\label{eq:S1}
\end{equation}
The last two terms in \eq{eq:S1} change only the chemical potential and add a constant energy shift, respectively, and thus do not modify our self-consistent solution. After Fourier transforming the total effective interaction and introducing the  basis of $\Psi_{\bm{k}}^\dagger = (c_{{\bm{k}} A \uparrow}^\dagger, c_{{\bm{k}} B \uparrow}^\dagger, c_{{\bm{k}} A \downarrow}^\dagger, c_{{\bm{k}} B \downarrow}^\dagger)$, we obtain the mean-field Hamiltonian in \eq{equ:7}.

The Hartree-Fock equations are solved  by self-consistently determining the order parameter $\delta m$
\begin{equation}
\begin{split}
\delta m &=\frac{1}{4 N}\langle n_{A \uparrow} -n_{B \uparrow} - n_{A \downarrow} +n_{B \downarrow}\rangle_\text{HF}\\\
&= \frac{1}{4 N} \sum_{\bm{k},\alpha} \psi^{\dagger}_{\alpha}(\bm{k})\vcenter{\hbox{$\begin{bmatrix}
    1 & 0 & 0 & 0 \\
    0 & -1 & 0 & 0 \\
    0 & 0 & -1 & 0 \\
    0 & 0 & 0 & 1 \\
\end{bmatrix}$}} \psi_{\alpha}(\bm{k}) \cdot f(\epsilon_{\bm{k},\alpha} - \mu)
\end{split}
\label{eq:Self-consist}
\end{equation}
and the chemical potential $\mu$ which fixes the total density
\begin{equation}
\begin{split}
    n &=\frac{1}{2 N}\langle n_{A \uparrow} + n_{B \uparrow} + n_{A \downarrow} +n_{B \downarrow}\rangle_\text{HF}\\
&= \frac{1}{2 N} \sum_{\bm{k},\alpha} 
f(\epsilon_{\bm{k},\alpha} - \mu).
\end{split}
\label{eq:Self-consistmu}
\end{equation}
Here, $f(\epsilon_{\bm{k},\alpha} - \mu)$ is the Fermi distribution, $\epsilon_{\bm{k},\alpha}$, $\psi_{\alpha}(\bm{k})$ are the eigenvalues and eigenvectors of  the matrix $H(\bm{k}) = \text{diag} (H_{\uparrow}(\bm{k}),H_{\downarrow}(\bm{k}))$, $\alpha$ is the band index running from one to four, and $N$ is the total number of unit cells. 

To self-consistently determine the order parameter $\delta m$ and the chemical potential $\mu$, we take a $200 \times 200$ square grid of momentum points in the magnetic Brillouin zone. For each momentum we evaluate the matrix $H(\bm{k}) = \text{diag} (H_{\uparrow}(\bm{k}),H_{\downarrow}(\bm{k}))$ and calculate the eigenvalues $\epsilon_{\alpha}(\bm{k})$ and eigenvectors $\psi_{\alpha}(\bm{k})$ of $H(\bm{k})$.

We solve the self-consistent equation by iteration. In each cycle of iteration we take an input value of $\delta m_\text{in}$ and calculate the resulting output $\delta m_\text{out}$ from \eq{eq:Self-consist}. In each iteration step we determine the chemical potential $\mu$ by fixing number of particles using \eq{eq:Self-consistmu}. For the next iteration loop we mix in 40\% of the previous solution to improve convergence.  We continue to run the iteration until the value of $\delta m$ converges to $|\delta m_\text{out} -\delta m_\text{in}| < 10^{-6}$.

\subsection{Tight-Binding Parameters}

\begin{figure}
	\centering
	\includegraphics[width=1\linewidth]{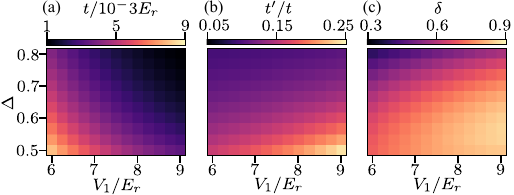}
	\caption{\textbf{Tight-binding parameters.} We compute the tight-binding parameters (a) nearest neighbor hopping $t/E_r$, (b) diagonal hopping $t'/t$, and (c) diagonal hopping anisotropy $\delta$, as a function of the optical lattice parameters $V_1$ and $\Delta$ for fixed $V_0 = 4E_r$.  }
	\label{fig:parameter}
\end{figure}

By varying the optical lattice potential characterized by $V_0, V_1, \Delta$, the parameters of the tight-binding model can be adjusted. Here, we evaluate the tight-binding parameters $t$, $t'$, and $\delta$ for fixed value of $V_0 = 4 E_r$ as a function of $V_1$ and $\Delta$; see Fig.~\ref{fig:parameter}. We find for a large regime of optical lattice parameters, favorable tight-binding parameters for stabilizing altermagnetism. 

\begin{figure*}
  \centering
\includegraphics{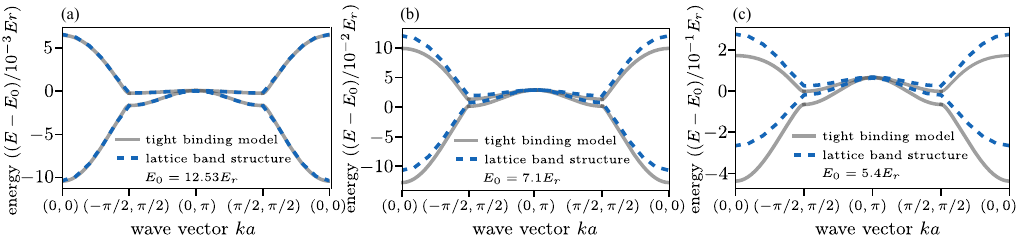}
    \caption{\textbf{Tuning the optical lattice potential.} Comparison of the numerically evaluated lattice band structure to the tight-binding band structure discussed in the main text for various depth of the optical lattice potential, characterized by $V_0$ and $V_1$. Data evaluated for (a) $V_0 = 4 E_r, V_1/V_0 = 2, \Delta = 0.7$, (b) $V_0 = 3 E_r, V_1/V_0 = 1, \Delta = 0.7$, (c) $V_0 = 2 E_r, V_1/V_0 = 1, \Delta = 0.7$.}
\label{fig:band}
\end{figure*}

\begin{figure*}
  \centering
\includegraphics{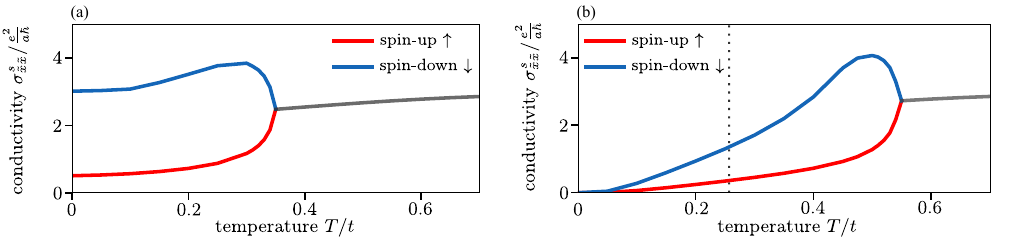}
    \caption{\textbf{Finite temperature spin conductivity.} Spin conductivity $\sigma^s_{\tilde{x}\tilde{x}}$ as a function of temperature for both spin components. Data evaluated for (a) AMM: $t'/t = 0.4, \delta = 0.5,U/t = 3.4$. (b) AMI: $t'/t = 0.4, \delta = 0.5, U/t = 4$, characterized by a gap of $0.26t$ (dashed line) }
    \label{fig:cond}
  \end{figure*}

\subsection{Tuning the optical lattice potential}

In the main text, we consider deep optical lattices, captured by a simple tight-binding model Eq.~(\ref{eq:Hamiltonian}) with a hopping amplitude $t$ on the order of $10^{-3} E_r$ ($E_r$ is the recoil energy). For experimental realizations larger values of the absolute hopping energy scale may be advantageous, for example to obtain homogeneous parameters throughout the system and to achieve comparatively low temperatures. 
Here, we numerically compute the band structure for various optical lattice depths; see Fig.~\ref{fig:band}. Although the simple tight-binding model does not quantitatively capture the band structure  anymore, we observe that the band structure still has the same symmetries and  anisotropic behavior as in the deep lattice. Thus, also in these lattices the altermagnetic phase is expected to be stabilized in a regime with experimentally favorable energy scales, even though the effective model requires the inclusion of longer ranged hoppings.

\subsection{Calculation of Conductivity Tensor}

The conductivity tensor is a $2 \times 2$ matrix for each spin
\begin{equation}
\begin{split}
  \sigma^s = \vcenter{\hbox{$\begin{bmatrix}
    \sigma^s_{\tilde x \tilde x} & \sigma^s_{\tilde x \tilde y} \\
    \sigma^s_{\tilde y \tilde x} & \sigma^s_{\tilde y \tilde y}\\
\end{bmatrix}$}}  
\end{split}
\label{eq:B1}
\end{equation}

We determine the matrix elements of the DC conductivity $\sigma_{\alpha \beta}^s(\omega \to 0)$ from the Kubo formula~\cite{Kubo_1957, Crepieux_2001, Freimuth_2014}

\begin{align}
 \sigma_{\alpha\alpha}^s (\omega \to 0) &=\frac{\hbar}{4\pi V}  \int_{-\infty}^{\infty} d\epsilon \frac{df}{d\epsilon}  \text{Tr}\langle v_\alpha^s (G^+-G^-) v_\alpha^s (G^+-G^-)\rangle,
\label{eq:B3}
\end{align}
where, $G^{\pm} (\epsilon)= [\epsilon - H \pm i\Gamma]^{-1}$ is the retarded/advanced Green's function and $\bm{v}^s$ is the velocity defined as, 
\begin{equation}
\begin{split}
 \bm{v}^s(\bm{k}) = \frac{1}{\hbar}\nabla_{\bm{k}} H_s(\bm{k}),
\end{split}
\label{eq:B4}
\end{equation}
where $H_s(\bm{k})$ is given in \eq{equ:7}. The difference between the retarded and advanced Green's functions is $G^+ - G^- = -2i \pi \delta(\epsilon - H)$. For our numerical evaluation we replace the delta distribution by a Lorentzian $\pi \delta(\epsilon - H) = {\Gamma}/[{(\epsilon - H)^2+\Gamma^2}]$ with broadening $\Gamma = 0.02$.

From diagonalizing the Hartree-Fock Hamiltonian $H(\bm{k})$ in \eq{equ:7} with self-consistently determined order parameter $\delta m$ and chemical potential $\mu$, we obtain the eigenvalues $\epsilon_{\alpha}(\bm{k})$ and eigenvectors $\psi_{\alpha}(\bm{k})$. We express the conductivity tensor in this basis as:

\begin{widetext}
\begin{equation}
\begin{split}
 \sigma_{\alpha\alpha}^s  
 &= \frac{\hbar}{4\pi V} \int_{-\infty}^{\infty} d\epsilon \frac{df}{d\epsilon} \sum_{m,n,\bm{k}} \langle \psi_m(\bm{k}) | v_\alpha^s |\psi_n(\bm{k})\rangle (G^+-G^-)(\epsilon_n) \langle \psi_n(\textbf{k})| v_i |\psi_m(\bm{k})\rangle (G^+- G^-)(\epsilon_m) \\
 &= -\frac{\hbar}{\pi V} \int_{-\infty}^{\infty} d\epsilon \frac{df}{d\epsilon} \sum_{m,n,\bm{k}} |\langle \psi_m(\bm{k})| v_\alpha^s |\psi_n(\bm{k})\rangle|^2   \frac{\Gamma}{(\epsilon - \epsilon_n)^2+\Gamma^2} \frac{\Gamma}{(\epsilon - \epsilon_m)^2+\Gamma^2}.
\end{split}
\label{equ:conductivity}
\end{equation}
\end{widetext}
We calculate \eq{equ:conductivity} numerically at finite temperature. At zero temperature, we  further simplify the evaluation by using $\frac{df}{d\epsilon} = - \delta(\epsilon - \epsilon_F)$.

\subsection{Conductivity as a function of temperature}

We evaluate the conductivity as a function of temperature both in the AMM and the AMI using Eq.(\ref{equ:conductivity}); see Fig. \ref{fig:cond}. For the AMI, the conductivity starts off at zero and increases with an activated behavior. By contrast the AMM has a finite conductivity down to zero temperatures. Above a critical temperature $T_c$ the altermagnetic order is destroyed and the conductivity tensor becomes symmetric. We note that this is an artefact of them mean-field treatment, as the Mermin-Wagner theorem states that order is absent at finite temperature in two spatial dimensions. Nonetheless, for our system the correlation length grows quickly as the temperature is lowered. Thus finite size systems appear to be ordered, provided the temperature is sufficiently low.

\newpage
\twocolumngrid

\end{document}